\documentclass[a4paper]{jpconf}
\usepackage{graphicx}
\usepackage{amstext}
\usepackage{subfigure}
\usepackage{graphicx}% Include figure files
\usepackage{dcolumn}% Align table columns on decimal point
\usepackage{bm}% bold math
\begin{document}
\title{The magnetic moments of the proton and the antiproton}

\author{S. Ulmer$^{1}$, A. Mooser$^{2,3}$, K. Blaum$^{4}$, S. Braeuninger$^{4}$,  K. Franke$^{1,4}$, H. Kracke$^{2,3}$, C. Leiteritz$^{2}$, Y. Matsuda$^{5}$, H. Nagahama$^{5}$, C. Ospelkaus$^{6}$, C. C. Rodegheri$^{2,4}$, W. Quint$^{8}$,  G. Schneider$^{1,2}$, C. Smorra$^{1}$, S. Van Gorp$^{7}$, J. Walz$^{2,3}$, and Y. Yamazaki$^{7}$}
\address{$^1$ RIKEN, Ulmer Initiative Research Unit,  2-1 Hirosawa, Wako, Saitama 351-0198, Japan}
\address{$^2$ Institut f\"ur Physik, Johannes Gutenberg-Universit\"at D-55099 Mainz, Germany}
\address{$^3$ Helmholtz-Institut Mainz,  D-55099 Mainz, Germany}
\address{$^4$ Max-Planck-Institut f\"ur Kernphysik, Saupfercheckweg 1, D-69117 Heidelberg, Germany}
\address{$^5$ Graduate School of Arts and Sciences, University of Tokyo, Tokyo 153-8902, Japan}
\address{$^6$ Leibniz Universit\"at Hannover, D-30167 Hannover, Germany}
\address{$^7$ RIKEN, Atomic Physics Laboratory,  2-1 Hirosawa, Wako, Saitama, 351-0198, Japan}
\address{$^8$ GSI - Helmholtzzentrum f\"ur Schwerionenforschung, D-64291 Darmstadt, Germany}

\date{\today}

\ead{stefan.ulmer@cern.ch}

\begin{abstract}
Recent exciting progress in the preparation and manipulation of the motional quantum states of a single trapped proton enabled the first direct detection of the particle's spin state. Based on this success the proton magnetic moment $\mu_p$ was measured with ppm precision in a Penning trap with a superimposed magnetic field inhomogeneity. An improvement by an additional factor of 1000 in precision is possible by application of the so-called double Penning trap technique. In a recent paper we reported the first demonstration of this method with a single trapped proton, which is a major step towards the first direct high-precision measurement of $\mu_p$. The techniques required for the proton can be directly applied to measure the antiproton magnetic moment $\mu_{\bar{p}}$. An improvement in precision of $\mu_{\bar{p}}$ by more than three orders of magnitude becomes possible, which will provide one of the most sensitive tests of CPT invariance. To achieve this research goal we are currently setting up the Baryon Antibaryon Symmetry Experiment (BASE) at the antiproton decelerator (AD) of CERN.
\end{abstract}

The local relativistic quantum field theories involved in the Standard Model of particle physics are CPT invariant \cite{CPT}.
As a consequence, in collisions particles and their antimatter mirror images annihilate completely. This is in strong contradiction to the striking imbalance of matter and antimatter observed in our Universe, which inspires comparisons of the fundamental properties of particles and their antiparticles with highest precision \cite{Yasu}. Any measured difference would shed light on this contradiction \cite{Dine}.
\\
In BASE \cite{TDR,CS} we focus on the measurements of the magnetic moments of the proton $\mu_{{p}}$ and the antiproton $\mu_{\bar{p}}$ with ppb precision or better. We store a single (anti)proton in a Penning trap \cite{Brown} and measure the ratio of its cyclotron frequency $\nu_c = 1/(2\pi)\cdot(e/m_p)\cdot B$ and its spin precession (Larmor) frequency $\nu_L = g/2 \cdot \nu_c$. $B$ is the magnetic field of the Penning trap, $e/m$ the particle's charge-to-mass ratio and $g$ the so-called $g$-factor. The measured frequency ratio
\begin{eqnarray}
\frac{\nu_L}{\nu_c}=\frac{g}{2}=\frac{\mu_{p,\bar{p}}}{\mu_N}
\end{eqnarray}
gives the dimensionless magnetic moment $\mu_{p,\bar{p}}$ in units of the nuclear magneton $\mu_N=(e\hbar)/(2m)$.
The cyclotron frequency $\nu_c$ is measured via image current detection \cite{Wine}, while $\nu_L$ is determined by using the continuous Stern-Gerlach effect \cite{DehmeltCSG}. In this detection scheme the spin magnetic moment is coupled to the particle's axial oscillation frequency $\nu_z$, which is achieved by superimposing a so-called magnetic bottle $B_z=B_2\left(z^2-{\rho^2}/{2}\right)$ to the homogeneous magnetic field $B$ of the trap, where $B_2$ characterizes the strength of the magnetic inhomogeneity and $z, \rho$ are cylindrical coordinates. A spin quantum jump shifts $\nu_z$ by
\begin{eqnarray}
\Delta\nu_{z,SF}=\frac{\mu_{p,\bar{p}} B_2}{2\pi^2m_{p,\bar{p}}\nu_z}.
\end{eqnarray}
Therefore, from a sequence of alternating axial frequency measurements and resonantly induced spin quantum transitions, the spin flip probability as a function of the spin-flip drive frequency $\nu_{rf}$ is obtained, and thus the Larmor frequency $\nu_L$ \cite{Brown2}.
This principle has been applied with great success in the widely recognized measurements of the electron and the positron magnetic moments \cite{VanDyck} providing one of the most stringent tests of CPT symmetry. However, the (anti)proton magnetic moment is 658 times smaller than that of the electron, which constitutes a significant challenge. Typical axial frequencies in Penning traps are in the order of 1$\,$MHz  and the axial frequency shift induced by a spin flip is $\Delta\nu_\text{z,SF}\approx0.4\,\mu$Hz$\cdot B_2$. Thus, to resolve single proton spin flips we developed a Penning trap with $B_2\approx300\,000\,$T/m$^2$. With this apparatus we reported the first observation of proton spin-transitions \cite{UlmerPRL} and measured the particle's magnetic moment with a relative precision of 8.9$\,$ppm \cite{CCR}. By applying the same method, an independent research group reported a precision of 2.5$\,$ppm for the proton \cite{Jack1} and 4.4$\,$ppm for the antiproton \cite{Jack2}. All these experiments were carried out in a Penning trap with the superimposed magnetic bottle. However, this produces an unavoidable line broadening of the spin resonance, ultimately limiting experimental precision to the ppm level. To overcome this problem, a group at Mainz developed the double Penning trap technique \cite{häffner2003double}. In this elegant scheme the precision measurement of $\nu_c$ and $\nu_L$, and the detection of the spin state are separated to two traps, a precision trap (PT) with a homogeneous magnetic field, and an analysis trap (AT) with the strong magnetic bottle.
In the BASE experiment the magnetic field in the precision trap is about 75\,000 times more homogeneous than that in the analysis trap which reduces the width of the spin-line significantly, thus enabling high-precision measurements.  By applying this technique in measurements of the $g$-factor of the electron bound to highly charged ions experimental precisions below the ppb level were achieved  \cite{Hermanspahn,Hartmut,Verdu,Sven}. \\

Our double Penning trap is shown in Fig.\ 1.
\begin{figure}[htb]
        \centerline{\includegraphics[width=11cm,keepaspectratio]{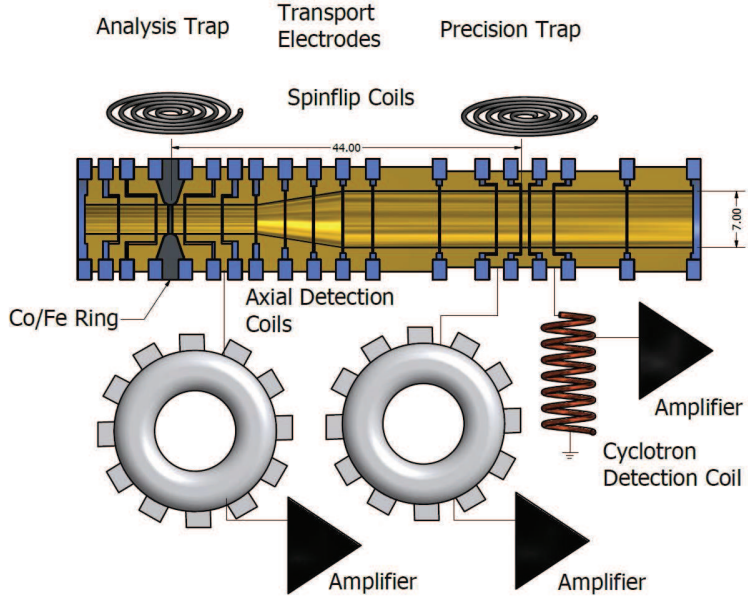}}
            \caption[FEP]{Double Penning trap setup used to measure the proton magnetic moment with high precision. Details are described in the text.}
            \label{fig:DoubleTrap}
\end{figure}
The cylindrical electrodes are made out of oxygen-free copper. They are gold-plated to prevent oxidation. The precision trap has an inner diameter of 7$\,$mm while that of the analysis trap is 3.6$\,$mm. This small diameter is required to produce the strong magnetic bottle by a central ring electrode which is made of ferromagnetic Co/Fe material. The two traps are connected by transport electrodes. Adequate voltage ramps applied to these electrodes are used to shuttle the (anti)proton adiabatically between the traps. Radio frequency (rf) drives applied to small coils mounted close to each trap produce oscillating magnetic fields which are used to induce spin quantum transitions \cite{Rabi1}. The whole setup is placed in a hermetically sealed vacuum chamber with a volume of about 1$\,$l, which is cooled to 4$\,$K. When cooled to cryogenic temperatures in this volume typically background pressures in the order of 10$^{-16}\,$mbar are achieved. This prevents particle loss due to charge exchange or annihilation with the background gas. \\
To measure the particle's eigenfrequencies highly sensitive detection systems are used \cite{ulmer2011Det}. They consist of coils with inductance $L$ and high quality factors $Q$, coupled to ultra low-noise amplifiers. Together with the trap capacitance $C_p$ they form parallel tuned circuits with resonance frequencies $\nu_\text{res}$. Once tuned to resonance with the particle's respective oscillation frequency $\nu_k$, image currents induced in the trap electrodes by the particle's oscillation lead to a voltage drop across the effective detection resistor $R_p=2\pi\nu_k L Q$. The signals are amplified by cryogenic ultra low-noise amplifiers and from a fast Fourier transform (FFT) the particle's resonance frequencies are obtained. By interaction with $R_p$ the particle is cooled resistively. The cooling time constant is
\begin{eqnarray}
\tau=\frac{m_{p,\bar{p}}}{R_p}\cdot\frac{D^2}{q_{p,\bar{p}}^2},
\end{eqnarray}
where $D$ is a trap specific length, 7.6$\,$mm in the PT and about 14$\,$mm in the AT. Once cooled to thermal equilibrium, the particle shorts the thermal Johnson noise of the detector \cite{UlmerPRL2}, resulting in a dip in the FFT-spectrum. In this case the particle's eigenfrequency is obtained from a best fit to the data. We obtain an axial frequency resolution of 60$\,$mHz in a measurement time of about 60$\,$s.\\
For the application of the double-trap technique we prepare a single proton in the AT and determine its spin state (see Fig.\ 2).
Next, we transport the single proton to the precision trap, and while a spin-flip drive is applied, the cyclotron frequency is measured.
In a last step the particle is transported back to the AT and the spin state is analyzed.
\begin{figure}[htb]
        \centerline{\includegraphics[width=13cm,keepaspectratio]{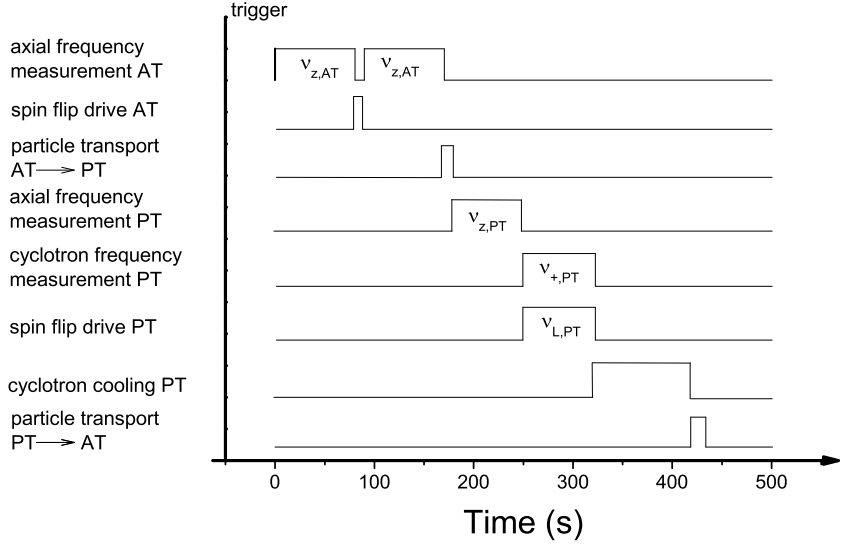}}
            \caption[FEP]{Illustration of a double Penning trap sequence. For details see text.}
            \label{fig:DoubleTrap}
\end{figure}
It is obvious that for the application of this technique ``single spin flip resolution'' is required \cite{MooserPRL}. This means, that the spin state of the proton has to be identified unambiguously. However, this is difficult.
The strong magnetic bottle $B_2$ couples in addition to the spin magnetic moment as well the magnetic moment of the orbital angular magnetic momentum of the radial modes to the axial oscillation frequency. Thus, small fluctuations of the magnetron and the cyclotron energy  cause axial frequency fluctuations.
A cyclotron quantum transition $\Delta n_+=\pm 1$ causes an axial frequency shift of $\Delta\nu_z=\pm63\,$mHz and a transition of the magnetron quantum number $\Delta n_-=\pm 1$ leads to $\Delta\nu_z=\pm54\,\mu$Hz. Thus, to clearly observe spin transitions it is crucial to keep axial frequency fluctuations induced by radial quantum transitions small in comparison to the frequency shift $\Delta\nu_\text{z,SF}=171\,$mHz.
\\
As an example we consider in the further discussion cyclotron quantum jumps. They are due to electric dipole transitions driven by spurious noise on the trap electrodes. The transition rate $\delta n_+/dt$ is proportional to the quantum number $n_{+}$,
as described in detail in \cite{MooserPRL}. We measured the axial frequency fluctuation as a function of the proton's average cyclotron quantum number. The details of the calibration and preparation procedure of $n_+$ are described in \cite{MooserPLB}. From this measurement we obtained the data shown in Fig.\ 3.
\begin{figure}[htb]
        \centerline{\includegraphics[width=10cm,keepaspectratio]{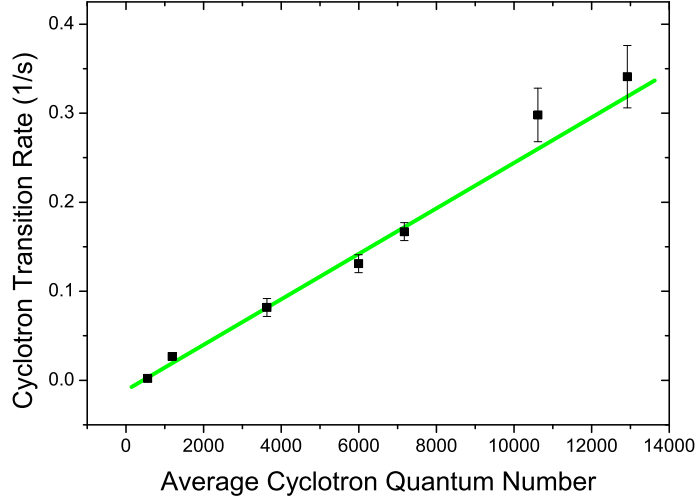}}
            \caption[FEP]{Cyclotron transition rate as a function of the average cyclotron quantum number. The filled squares are measured data, the solid line is a fit. Once the proton's cyclotron quantum state is prepared below $n_+\approx1500$, we achieve single spin flip resolution with a fidelity of about 75$\,\%$.}
            \label{fig:DoubleTrap}
\end{figure}
The heating-rate increases as a function of the average cyclotron quantum number, which confirms our semi-quantitative understanding of the axial frequency stability.
\\
To apply the double Penning trap technique we prepare the particle with an average cyclotron quantum number below $n_+\approx1500$. There, the cyclotron transition rate is below 0.03/s. This is achieved by resistive cooling of the modified cyclotron mode in the precision trap which takes about 2 hours.  Once the cyclotron quantum state is prepared, we achieve a spin flip fidelity of 75$\,\%$. This means, that three out of four detected spin transitions are correctly identified \cite{MooserPRL}. For a demonstration of the double-trap technique in the precision trap resonant spin flips are driven with a drive amplitude strong enough to saturate the spin transition. For a background measurement we applied the whole double-trap scheme, however no spin flip drive was applied in the PT. Results of these measurements are shown in Fig.\ 4. For the resonantly driven spin flips we obtain a spin flip probability of 50$\,\%$, while in the background measurement only 25$\,\%$ are obtained. This is due to the 75$\,\%$ spin flip fidelity in the AT. Under these conditions, when measuring a full Larmor resonance curve in the PT, the maximum to baseline ratio of the resulting spin line would be only about 25$\,\%$. This reduced ''contrast'' affects the achievable experimental precision in the $g$-factor resonance. A possible way to improve the contrast of the resonance is the application of a phase sensitive axial frequency measurement technique \cite{Stahl}.  By using this method we measured the axial frequency in the precision trap within a few seconds, achieving spin flip resolution \cite{UlmerPhD}. Once implemented in the analysis trap we expect to obtain spin flip fidelities of 100$\,\%$.\\
\begin{figure}[htb]
        \centerline{\includegraphics[width=10cm,keepaspectratio]{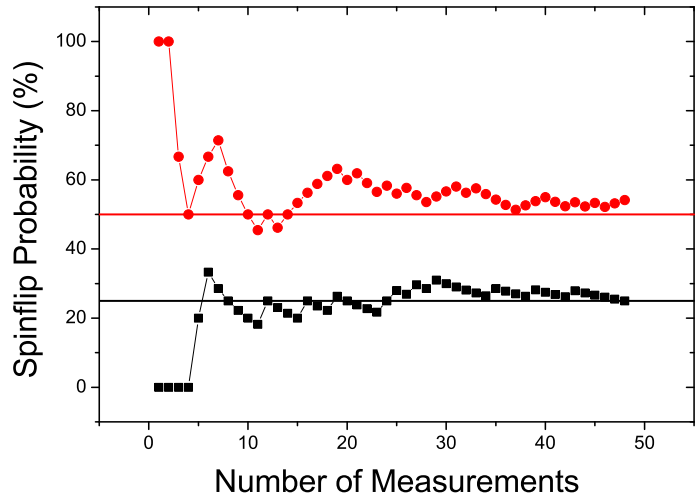}}
            \caption[FEP]{Spin flip probability as a function of measuring cycles. The filled circles represent data were spin flips were driven in the PT. The filled squares represent results of a background measurement where the double-trap sequence has been applied, however no spin flips were induced in the PT. When the spin flip drive was turned on a probability of 50$\,\%$ is measured, corresponding to saturation. When no spin flip drive is applied a probability of 25$\,\%$ is measured, which is due to the finite fidelity.}
            \label{fig:DoubleTrap}
\end{figure}
To apply our methods to the antiproton we are currently setting up the BASE experiment \cite{TDR,CS} at the antiproton decelerator of CERN \cite{AD1,AD2}. Essentially, the antiproton magnetic moment can be measured by applying exactly the same techniques as the ones described above. However loading of antiprotons into our Penning trap apparatus requires several modifications of the trap and the implementation of the experiment into the high energy physics infrastructure of CERN. A new transfer beamline for the 5.3$\,$MeV antiprotons provided by the AD has been designed and verified by the AD group. The shielded experimental zone where the apparatus will be installed is currently being constructed. In addition to the PT and the AT which are required for the $g$-factor measurement a so-called catching trap as well as a degrader structure will be included into our apparatus, which is shown in Fig.\ 5. The degrader structure is a sequence of thin stainless steel (SUS), Be and Al foils of variable thickness. Once the foil sequence is carefully chosen and the thickness of each foil is adjusted carefully, about 1 per-mille of the incident antiprotons are decelerated to energies below 15$\,$keV. Downstream to the degrader a specifically designed catching trap (CT) will be placed. 15$\,$kV catching pulses can be applied to the trap electrodes to catch the degraded antiprotons. These are standard methods which were developed and optimize in the late 1980ies \cite{JerryTrappedAntiprotons}. Once the particles are trapped, electron cooling \cite{ElectronCooling} and resistive cooling will be used to prepare the 100$\,\mu$eV particle energies, which are required for our high precision measurement. To become independent from accelerator shutdown and power cuts, our CT will be operated on battery based voltage sources. We plan to load several AD antiproton shots into this trap and techniques will be developed to prepare single antiprotons into the AT/PT cycle to perform the precision measurement of the antiproton magnetic moment $\mu_{\bar{p}}$. Together with the planned measurement of the ground state hyperfine splitting of antihydrogen \cite{Enomoto}, constraints on the antiproton sub-structure will be obtained.
\begin{figure}[htb]
        \centerline{\includegraphics[width=11cm,keepaspectratio]{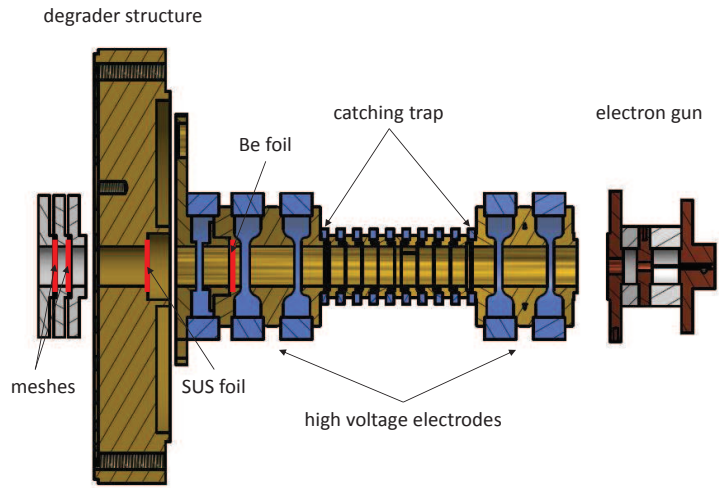}}
            \caption[FEP]{BASE degrader system and catching trap.}
            \label{fig:DoubleTrap}
\end{figure}
Besides the precision measurement of the antiproton magnetic moment we plan as well to improve the value of the proton to antiproton charge-to-mass ratio. This number has been measured in the late 1990ies and is known with a relative precision of about 90$\,$ppt \cite{JerryAntiproton}. However, in recent years advanced mass spectrometry techniques have been developed and mass measurements at the level of 5$\,$ppt were reported \cite{Rainville}. By applying these techniques the values of the proton to antiproton charge-to-mass ratio might be improved by another order of magnitude.
\\
In this paper the status of the BASE experiment was presented. By using a Penning trap with a strong superimposed magnetic inhomogeneity we detected spin flips with a single trapped proton and measured the particle's magnetic moment with a relative precision of 8.9$\,$ppm. With an improved apparatus we achieved single spin flip resolution. Based on this success we reported on the first demonstration of the double-trap method with a single trapped proton, which is a major step towards a direct high precision measurement of $\mu_p$ on the ppb-level. Currently we are setting up the BASE experiment at CERN to apply our techniques to a single trapped antiproton.
\\
\\
We would like to express our gratitude towards Lajos Bojtar, Tommy Eriksson, Francois Butin, Theodore Rutter and Stephan Maury, who are responsible for the construction of the BASE antiproton transfer line and the implementation of the new BASE zone into the infrastructure of the AD. We are also thankful towards all CERN working groups which contribute to BASE.\\
We acknowledge financial support of RIKEN Initiative Research Unit Program, the Max-Planck Society, the BMBF, the Helmholtz-Gemeinschaft, HGS-HIRE, and the EU (ERC Grant No. 290870-MEFUCO).

\end{document}